# Gamma ray vortices from nonlinear inverse Compton scattering of circularly polarized light

Yoshitaka Taira[1,2], Takehito Hayakawa[3,4], Masahiro Katoh[5*]

[1]*Research Institute for Measurement and Analytical Instrumentation, National Metrology Institute of Japan, National Institute of Advanced Industrial Science and Technology (AIST), Tsukuba Central 2, 1-1-1 Umezono, Tsukuba, Ibaraki 305-8568 Japan*

[2]*Department of Physics and Astronomy, Mississippi State University, 355 Lee Blvd., 125 Hilbun Hall, Mississippi State, MS 39762 USA*

[3]*National Institutes of Quantum and Radiological Science and Technology, Shitakata 2-4, Tokai, Ibaraki 319-1106 Japan*

[4]*National Astronomical Observatory of Japan, 2-21-1 Osawa, Mitaka, Tokyo 181-8588 Japan*

[5]*Institute for Molecular Science, National Institutes of Natural Sciences/School of Physical Sciences, SOKENDAI (The Graduate University for Advanced Studies), Myodaiji-cho, Okazaki 444-8585 Japan*

*e-mail: mkatoh@ims.ac.jp

Inverse Compton scattering (ICS) is an elemental radiation process that produces high-energy photons both in nature and in the laboratory[1]. Non-linear ICS is a process in which multiple photons are converted to a single high-energy photon[2]. Here, we theoretically show that the photon produced by non-linear ICS of circularly polarized photons is a vortex, which means that it possesses a helical wave front and carries orbital angular momentum. Our work explains a recent experimental result regarding non-linear Compton scattering that clearly shows an annular intensity distribution as a remarkable feature of a vortex beam. Our work implies that gamma ray vortices should be produced in various situations in astrophysics in which high-energy electrons and intense circularly polarized light fields coexist. They should play a critical role in stellar nucleosynthesis. Non-linear ICS is the most promising radiation process for realizing a gamma ray vortex source based on currently available laser and accelerator technologies, which would be an indispensable tool for exploring gamma ray vortex science.

ICS is a process by which a high-energy electron scatters a low energy photon and converts it to a high-energy one. It is one of the elementary radiation processes that can produce X-rays or gamma rays in nature or in the laboratory. In fact, ICS beams are available at Duke University[3], NewSUBARU[4], and UVSOR[5] for the study of nuclear physics and various applications. A new facility, ELI-NP, will further explore this research field[6]. Progress in laser physics has enabled us to generate non-linear ICS (NICS), which is a multi-photon process that occurs when the light field is sufficiently intense. The classical picture of NICS[7-9] is as follows. In the electron rest frame, the electron oscillates in the incoming electromagnetic field. When the light is circularly polarized, the electron motion is also circular. When the field intensity is sufficiently high, such that the laser strength parameter, $a_0$, is almost unity, the electron motion becomes relativistic and radiates not only at the fundamental frequency of the circular motion but also at the harmonic frequencies.

From a quantum mechanics perspective, $n$ photons (n>1) are incoming, and only one photon is outgoing. When the in-coming photons are circularly polarized, each photon brings $\hbar$ angular momentum (AM) as spin angular momentum (SAM) and, in total, they bring $n\hbar$ AM to the system. Here, $\hbar$ is the Planck constant divided by $2\pi$. The outgoing single photon can carry at most $\hbar$ AM as SAM. How should it manage the remainder $(n-1)\hbar$ AM? As far as we know, no one has addressed this problem. Our answer is that the photon should carry the remainder as orbital angular momentum (OAM)[10]. In this work, we show theoretically for the first time that the photon produced by NICS of circularly polarized light has a helical wave front and carries OAM.

We shall treat NICS in the context of classical electrodynamics. The Fourier component of the electric field emitted by a single electron can be calculated from the Lienard–Wiechert potentials[11], considering the electron motion in the circularly polarized light field as described in the *Methods* section. The final form under the paraxial approximation can be expressed by the complex orthogonal unit vectors $\vec{e}_{\pm} \equiv (\vec{e}_x \pm i\vec{e}_y)/\sqrt{2}$, corresponding to the circular polarizations of positive and negative helicities, respectively, as follows:

$$\vec{E}(\omega) = \sum_{n=1}^{\infty} \left\{ \begin{array}{l} \dfrac{i}{\sqrt{2}}(C_\theta + C_\phi)\dfrac{\exp i\{kz+(n-1)\phi\}}{z}\vec{e}_+ \\ +\dfrac{i}{\sqrt{2}}(C_\theta - C_\phi)\dfrac{\exp i\{kz+(n+1)\phi\}}{z}\vec{e}_- \end{array} \right\}. \qquad (1)$$

Here, θ is the polar angle to the z-axis, φ is the azimuthal angle around the z-axis, z is the distance from the origin of the electron orbit to the observation plane, $C_\theta(\theta)$ and $C_\phi(\theta)$ are real functions of $\theta$ and are described in detail in the *Methods* section, and n is the harmonic number related to the wave number, ω, by the following equation:

$$\omega = \frac{8n\gamma_0^2 \omega_0}{2\gamma_0^2 \theta^2 + 2 + a_0^2}, \quad (2)$$

where $\gamma_0$ is the Lorentz factor of the electron and $\omega_0$ is the wave number of the incident photon.

Eq. (1) shows that the electric field is an elliptically polarized wave, which can be decomposed into circularly polarized components of positive and negative helicities. The higher harmonics ($n \geq 2$) of the positive helicity component possesses a helical phase structure represented by the phase term $\exp(i(n-1)\phi)$, which implies that the photons carry $(n-1)\hbar$ OAM, but the fundamental harmonic (n=1) does not. On the other hand, all harmonics of the negative-helicity component carry $(n+1)\hbar$ OAM. This finding may be interpreted as follows: each photon carries AM of $n\hbar$ in total and, according to its polarization state, carries $\pm\hbar$ as SAM and the remainder as $(n\mp1)\hbar$ OAM. We have noted that such polarization properties and vortex phase structures can be observed more commonly in the radiation fields emitted by relativistic electrons in circular motion; this topic is discussed in a separate paper.

The polarization state of the field may be represented by the Stokes parameter, $S_3/S_0$, which is expressed as follows[11]:

$$\frac{S_3}{S_0} = \frac{2 C_\theta C_\phi \cos\theta}{C_\theta^2 \cos^2\theta + C_\phi^2}. \quad (3)$$

The distributions of the Stokes parameter are shown in Fig. 1. The calculation parameters are given in the caption. Around the z-axis, the polarization is circular with positive helicity, which coincides with the electron motion. The photons in this region carry $(n-1)\hbar$ OAM. The degree of circular polarization decreases as the scattering angle increases. The polarization changes from circular to linear polarization at a large scattering angle. The polarization again changes to circular polarization with negative helicity at a greater scattering angle, and the photons carry $(n+1)\hbar$ OAM.

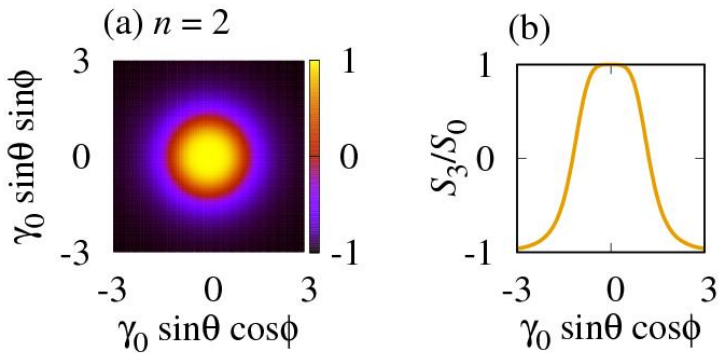

**Figure 1. Calculated Stokes parameters of circularly polarized gamma rays.** The (a) spatial and (b) line distributions of the degree of circular polarization in the x–y plane calculated by Eq. (3). The parameters are $\gamma_0 = 2000$, $a_0 = 1.0$, $n = 2$, and $\lambda_0 = 1.0$ μm.

The radiation energy is expressed as follows[11]:

$$\frac{d^2 I}{d\omega d\Omega} = 2\varepsilon_0 c R^2 |\vec{E}|^2. \quad (4)$$

Here, $\Omega$ is the solid angle. The explanation of the other symbols is described in the *Methods* section. The number of photons scattered to the polar angle between $\theta_1$ and $\theta_2$ can be approximately calculated as follows:

$$N = \frac{N_e}{\hbar} \int_{\theta_1}^{\theta_2} d\theta \int_0^{2\pi} d\phi \frac{d^2 I}{d\omega d\Omega} \sin\theta, \quad (5)$$

where $N_e$ is the number of electrons interacting with the laser per second.

The intensity distribution along the x-axis for each harmonic number is shown in Fig. 2. Most photons are concentrated on the z-axis, where the polarization state is predominantly circular with positive helicity, and the OAM value is $(n-1)\hbar$. An annular shape and zero intensity at the z-axis appear only for the harmonics but not for the fundamental frequency, as expected.

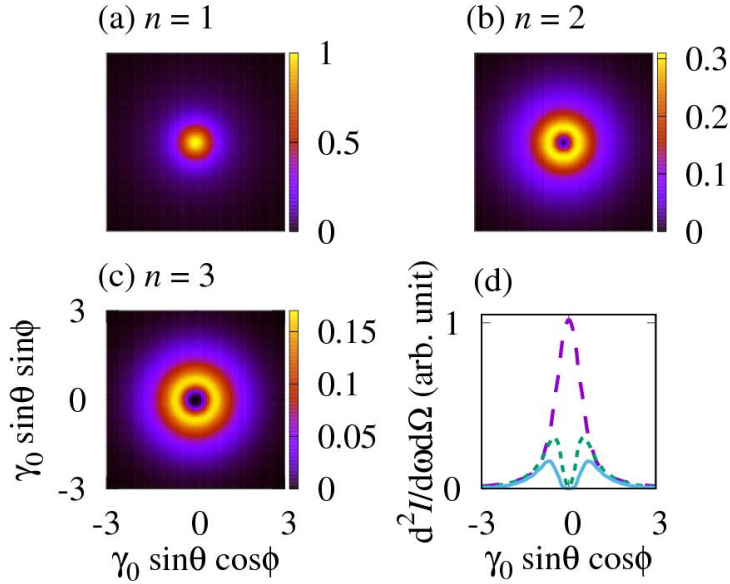

**Figure 2. Calculated spatial distribution of gamma rays.** Spatial intensity distributions of the (a) first, (b) second, and (c) third harmonics, calculated by Eq. (4) and normalized by the maximum value of the first harmonic. (d) Line intensity distribution along the x-axis for each harmonic number. The lines indicate the following: long dashed line, n = 1; short dashed line, n = 2; and solid line, n = 3. The other parameters are the same as in Figure 1.

We estimate the expected number of photons from a laboratory gamma ray vortex source based on NICS. The results are shown in Fig. 3, and the parameters described in the caption are readily available at modern mid-scale accelerator facilities. We found that in the scattering angle ranges $\theta < 0.6/\gamma_0$ and $2.40/\gamma_0 < \theta < 2.47/\gamma_0$, more than 90% of photons are circularly polarized with positive or negative helicity, respectively. The numbers of photons for both polarizations were obtained by integrating in these ranges. For example, one can obtain $6 \times 10^{11}$ photons/s as the fundamental component with positive helicity with an energy of 11–13 MeV, which does not carry OAM. In contrast, the second harmonic contains $2 \times 10^{11}$ photons/s with the energy of 21–26 MeV and carries $\hbar$ OAM. At the large angle of $2.40/\gamma_0 < \theta$, photons dominantly possess negative helicity, and all harmonics carry $(n+1)\hbar$ OAM; however, the number of photons is smaller by one order of magnitude than the ones with positive helicity.

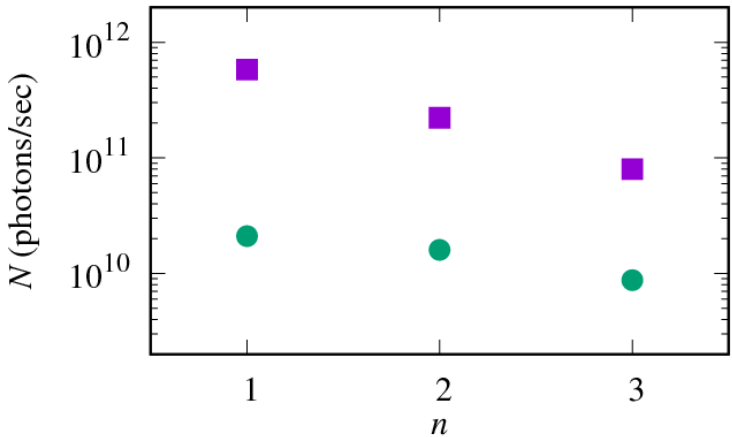

**Figure 3 Number of photons per unit time and solid angle at each harmonic calculated using Eq. (5).** Integration range of the scattering angle of each symbol as follows: square, $\theta_1 = 0.003/\gamma_0$ and $\theta_2 = 0.6/\gamma_0$; circle, $\theta_1 = 2.40/\gamma_0$ and $\theta_2 = 2.47/\gamma_0$. These regions correspond to the more than 90% of circular polarization with positive helicity carrying $(n-1)\hbar$ OAM and more than 90% of circular polarization with negative helicity carrying $(n+1)\hbar$ OAM, respectively. The calculation parameters are as follows: $\gamma_0 = 2000$, $\lambda_0 = 1.0$ μm, $a_0 = 1.0$, $N_e = 10^9$ electrons/sec, and $N_0 = 500$.

NICS using a relativistic electron beam and a laser has been experimentally investigated at accelerator facilities[12-14]. Recently, NICS was demonstrated for a circularly polarized laser with a strength parameter of $a_0 = 0.6$. The results showed an annular shape of the intensity distribution for the second harmonics of scattered X-rays at an energy of 13 keV[14]. Using Eq. (4), we can reproduce the experimental results, as shown in Fig. 4a. As a further verification, we suggest the measurement of a diffraction image produced by the interference between a wave field produced by a wire of a few μm and the vortex wave field[15], although to apply this method to NICS, the X-rays would need to be highly collimated. We expect to find a singularity in the diffraction pattern as shown in Fig. 4b and 4c.

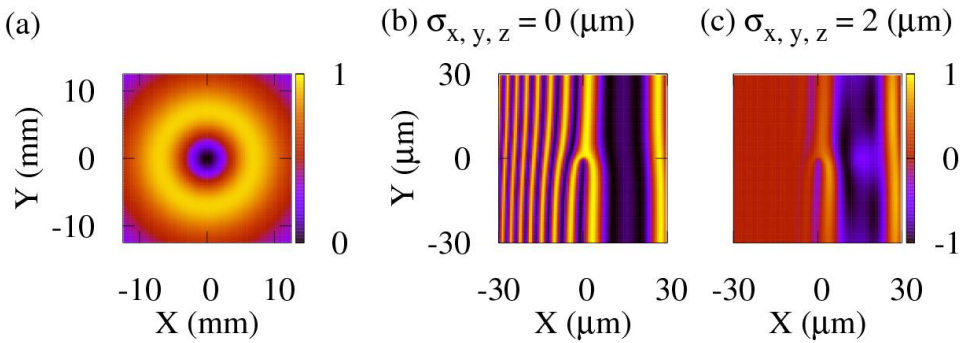

**Figure 4 Calculated spatial distribution of X-rays and interference pattern in the case of Brookhaven National Laboratory.** (a) The spatial distribution of the second harmonic calculated by Eq. (4). The parameters are $\gamma_0 = 128$, $a_0 = 0.6$, n = 2, and $\lambda_0 = 10.6$ μm. (b) and (c) The interference pattern between the X-ray vortex and diffracted X-ray from the wire. The parameters are n = 2, $\lambda = 0.095$ nm, $x_{off} = 15$ μm, $z_{off} = 0.5$ m, z = 1.85 m. Here, X-rays are assumed to be collimated by an infinitesimally small aperture (b) and a 2 μm one (c).

Gamma rays with energies of several MeV can interact with atomic nuclei with the unique feature that they are able to selectively excite states with specific spin and parity. In the photon-induced reaction in the $J^\pi = 0^+$ ground state, the excited state with $J^\pi = 1^-$, $1^+$, and $2^+$ can be selectively populated by an electric-dipole (E1), magnetic-dipole (M1), or electric-quadrupole (E2) transition. The excited state subsequently decays by the emission of a gamma ray and/or a particle such as a neutron. The gamma ray vortices carrying total angular momentum larger than 1 $\hbar$ will open up a new field of nuclear physics, as excited states can be populated by high-order transitions such as E3, E4, M2, and so on.

Photons in an energy range of several MeV play an important role in stellar nucleosynthesis, such as photodisintegration reactions in a supernova explosion (γ-process), in which some rare isotopes are produced by successive photodisintegration reactions[16]. If vortex photons interact with nuclei, the photon-induced reaction cross-sections may be drastically changed, potentially changing the isotopic abundances. In addition, photon-induced nuclear reactions strongly affect other explosive nucleosyntheses such as the r-process because photodisintegration reactions can destroy newly synthesized nuclides. Here, we emphasize that high-energy gamma ray vortices should be generated in stellar environments in which high-energy electrons and intense circularly polarized electromagnetic fields coexist[17-19], for example, the vicinity of magnetized neutron stars[20] or magnetohydrodynamic (MHD) jet supernova explosions[21]. Therefore, the NICS gamma rays from circularly polarized photons should play a critical role in stellar nucleosynthesis, and in the near future, a new facility generating gamma ray vortices will advance the study of stellar nucleosynthesis and the nuclear physics correlated with the vortex photons.

*Methods*

The Fourier component of the electric field emitted by a single electron can be calculated from the Lienard–Wiechert potentials[11]:

$$\vec{E}(\omega) = -i\sqrt{\frac{e^2 k^2}{32\pi^3 \varepsilon_0^2}} \frac{\exp(ikR)}{R} \int_{-\infty}^{\infty} dt \{\vec{n} \times (\vec{n} \times \vec{\beta})\} \exp\left\{i\omega\left(t - \frac{\vec{n} \cdot \vec{r}(t)}{c}\right)\right\}, \quad (6)$$

where $e$ is the elementary charge, $\omega$ and $k = \omega/c$ are the angular frequency and the wave number, $\varepsilon_0$ is the permittivity of vacuum, $R$ is the distance from the origin of the electron orbit to the observation point, $t$ is the emitter time, $\vec{n}$ is a unit vector pointing from the origin to the observation point, $\vec{\beta} = \vec{v}/c$ is the electron velocity, c is the speed of light, and $\vec{r} = (x, y, z)$ is the electron orbit. The electron orbit in the intensely circularly polarized laser is helical and given by[7]

$$\begin{aligned} x(\eta) &= x_0 + (r_1/\sqrt{2})\sin k_0 \eta \\ y(\eta) &= y_0 \mp (r_1/\sqrt{2})\cos k_0 \eta , \\ z(\eta) &= z_0 + \beta_1 \eta \end{aligned} \quad (7)$$

where $(x_0, y_0, z_0)$ are related to the initial position of the electron and, if we choose the origin of the coordinate at the centre of the electron motion, are all zero. Here, $r_1 = a_0/(h_0 k_0)$, $\beta_1 = (1-1/M_0)/2$, $k_0 =$ $2\pi/\lambda_0$, and $\eta = z + ct$. Here, $M_0 = h_0^2/(1 + a_0^2/2)$, $h_0 = \gamma_0(1+\beta_0)$, where $\beta_0$ is the initial normalized electron velocity, $a_0$ is the laser strength parameter described as $a_0 = 0.85 \times 10^{-9} \lambda_0(\mu m) I_0^{1/2}(W/cm^2)$, and $I_0$ is the intensity of the laser. Eq. (6) corresponds to the case in which the electron moves counterclockwise when the observer is facing the oncoming electron, which we call *positive helicity*.

The electric field calculated from Eqs. (6) and (7) is expressed in terms of complex orthogonal unit vectors as follows:

$$E(\omega) = \sum_{n=1}^{\infty} \left\{ \begin{array}{l} \dfrac{i}{\sqrt{2}}(C_\theta \cos\theta + C_\phi) \dfrac{\exp i\{\psi_0 + kR + (n-1)\phi\}}{R} e_+ \\ + \dfrac{i}{\sqrt{2}}(C_\theta \cos\theta - C_\phi) \dfrac{\exp i\{\psi_0 + kR + (n+1)\phi\}}{R} e_- \\ - iC_\theta \sin\theta \dfrac{\exp i\{\psi_0 + kR + n\phi\}}{R} e_z \end{array} \right\}. \qquad (8)$$

Here,

$$C_\theta = \sqrt{\dfrac{e^2 k^2 \lambda_0^2 N_0^2}{32\pi^3 \varepsilon_0^2 c^2}} \left( \dfrac{\sin \bar{k}\eta_0}{\bar{k}\eta_0} \right) \left( \dfrac{nk_0 \cos\theta}{k \sin\theta} - \beta_1 \sin\theta \right) J_n(p), \quad (9)$$

$$C_\phi = \sqrt{\dfrac{e^2 k^2 \lambda_0^2 N_0^2}{32\pi^3 \varepsilon_0^2 c^2}} \left( \dfrac{\sin \bar{k}\eta_0}{\bar{k}\eta_0} \right) \dfrac{a_0}{\sqrt{2}h_0} J_n'(p), \quad (10)$$

$$\psi_0 = -k\{x_0 \sin\theta \cos\phi + y_0 \sin\theta \sin\phi + z_0(1 + \cos\theta)\} \quad (11)$$

$$\bar{k} = k\{1 - \beta_1(1 + \cos\theta)\} - nk_0, \quad (12)$$

$J_n(p)$ and $J_n'(p)$ are the Bessel function of the first kind and its derivative, $p = (kr_1\sin\theta)/\sqrt{2}$, $\eta_0 = N_0\lambda_0/2$ ($N_0$ is the number of the periods of the laser field interacting with the electron, and $\lambda_0$ is its wavelength). Under the paraxial approximation, Eq. (8) is reduced to Eq. (1).

The diffraction of the gamma ray vortex by a wire is calculated as follows. A modulation of the intensity distribution given by the cosine of the interference term is expressed as[15]

$$\cos\left\{(n-1)\phi + kR + \psi_0(\theta,\phi) - (n-1)\phi_{off} - k\sqrt{(x - x_{off})^2 + (z - z_{off})^2} - \psi_0(\theta_{off}, \phi_{off})\right\}, \quad (13)$$

where $z$ is the distance between the interaction point and the detector plane, $x_{off}$ and $z_{off}$ are the offset coordinates of the wire with respect to the interaction point, $\theta_{off} = \tan^{-1}(\sqrt{(x-x_{off})^2 + y^2}/(z-z_{off}))$, and $\phi_{off} = \tan^{-1}(y/x_{off})$. Figure 4 (b) and (c) show the modulation of the intensity distribution produced by the cosine term given by Eq. (13). The wavelength of the X-ray, $\lambda = 2\pi/k$, is calculated to be 0.095 nm. The detector plane is set at z = 1.85 m. The wire is set at the position inside the vacuum pipe of $x_{off}$ = 15 μm and $z_{off}$ = 500 mm. We defined the initial position of the electron, $x_0$, $y_0$, and $z_0$, as following a Gaussian distribution with the width of the root mean square of $\sigma_x$, $\sigma_y$, and $\sigma_z$. Figure 4 (b) and (c) show the cases of $\sigma_x = \sigma_y = \sigma_z = 0$ μm and $\sigma_x = \sigma_y = \sigma_z = 2$ μm, respectively. One can see that the X-rays must to be collimated to a few μm scale to apply this method to NICS X-rays/gamma rays if the real beam size of the electron exceeds these values.

*Acknowledgements*

We would like to thank Prof M. Hosaka, Dr T. Kaneyasu, and Dr Joseph Grames for their helpful discussions. This work was supported by JSPS Postdoctoral Fellowships for Research Abroad, and a part of this work was supported by JSPS KAKENHI Grant Number 26286081.